\documentclass{ws-procs961x669}

\begin{document}


\title{Proper-time flow equation and non-local truncations in quantum gravity}

\author{E.M. Glaviano$^*$ and A. Bonanno $^{**}$}

\address{Dipartimento di Fisica e Astronomia “Ettore Majorana”, Università di Catania, Via S. Sofia 64, 95123, Catania, Italy\\ INAF, Osservatorio Astrofisico di Catania, via S.Sofia 78, I-95123 Catania, Italy \\
E-mail: * emiliano.glaviano@inaf.it\\ ** alfio.bonanno@inaf.it}

\begin{abstract}
We study the flow of the non-local truncation in quantum gravity and we focus in particular on the Polyakov effective action for a non-minimally coupled scalar field on a two dimensional curved space. We show that it is possible to explicitly integrate the flow of all the local and non-local operator terms up to $k=0$ and recover effective action without the integration of the conformal anomaly.
\end{abstract}


\bodymatter

\section{Introduction}
Functional flow equations based on proper-time (PT) regulators have garnered significant interest in recent years due to their effectiveness in various non-perturbative contexts. These include, for example, the study of the ordered phase in scalar field theories \cite{Bonanno:2019ukb} and investigations into non-perturbative quantum gravity \cite{Bonanno:2004sy}. In the latter case, the discovery of a non-trivial fixed point within the Einstein-Hilbert truncation has sparked interest in whether such a fixed point persists in more sophisticated truncations, particularly those involving an infinite number of terms and non-local invariants. A key question is whether these flow equations can accurately describe the evolution of non-local terms. Reproducing well-known non-local effective actions at $k=0$ is an important test of this approach.

In the simple case of a minimally coupled scalar field theory on a curved two-dimensional manifold, the effective action is exactly known as the Polyakov action. In this contribution we consider a theory of a non-minimally coupled scalar field on a two-dimensional curved space and use the non-local heat kernel trace expansion to demonstrate how PT flow equations can reproduce the Polyakov action at $k=0$. For the explicit calculations we will work in Euclidean signature.

\section{The functional renormalization group in a nutshell}
Let us review in a nutshell the functional renormalization group. The starting point is the Wilson’s idea about field theories. According to this approach, at some UV scale $\Lambda$ a theory will be described by a bare action that can be expressed as an expansion of a set of operators $O[\phi]$, which depend on the fields and its derivatives 
\begin{equation}
S_\Lambda\left[\phi\right]=\sum_{n=1}^{\infty}{g_nO_n\left[\phi\right]}
\end{equation}
where $g_n$ are the coupling constants of the theory. If we are interested in the description of physics at some momentum scale $k$ smaller than $\Lambda$, this comes up “integrating out” all fluctuations of the fields with momenta larger than $k$. This means that the microscopic degrees of freedom $\phi$ in the path integral are replaced by effective degrees of freedom $\psi$ obtained from a coarse graining procedure at scale $k$
\begin{equation}
e^{-S_k\left[\psi\right]}=\int{D\phi P_k[\phi,\psi]e^{-S_\Lambda[\phi]}}
\end{equation}
The coarse graining is implemented with a constraint operator $P_k[\phi,\psi]$ and the action $S_k$ is called a "Wilsonian action". Performing the integral shows that the effect of integrating out the microscopic degrees of freedom is that the coupling constants in the bare action acquire a scale dependence
\begin{equation}
S_k\left[\psi\right]=\sum_{n=1}^{\infty}{g_n\left(k\right)O_n\left[\psi\right]} 
\end{equation}
In this way a non-perturbative flow arises. In particular, these runnings stem from a set of differential equations called the functional renormalization group equations 
\begin{equation}
k\partial_kg_n\left(k\right)=\beta_n\left(g_1,g_2,\ldots,;k\right)
\end{equation}
the right-hand side of this equation is usually called the beta functions of the theory.

The Wilsonian action has a technical problem. In field theory we are interested in the computation of the one particle irreducible connected correlation functions but the Wilsonian action is not a functional generator for them. To overcome this difficulty the effective action 
\begin{equation}
\mathrm{\Gamma}\left[\phi\right]=\sum_{n=1}^{\infty}{g_n \bar O_n\left[\phi\right]}
\end{equation}
has to be considered. The Wilson’s idea then can be implemented replacing it with an “effective average action”. With the same logic as before, this means that the quantum fluctuations turn the coupling constants into running coupling constants 
\begin{equation}
\mathrm{\Gamma}_k\left[\phi\right]=\sum_{n=1}^{\infty}{g_n\left(k\right)\bar O_n\left[\phi\right]}
\end{equation}
In other terms, if no microscopic degree of freedom is integrated out the bare action at the UV scale $\Lambda$ must be obtained, whereas if we integrate out the microscopic degrees of freedom up to $k=0$ we must be left with the quantum effective action. Therefore, the effective average action interpolates between the quantum effective action and the bare action \cite{book}.

\section{The functional renormalizaton group and the proper time formalism}
The main task of functional renormalization group is to find a set of differential equations for the Wilsonian coupling constants, integrate them and reach $k=0$ to get the full effective action. To derive the proper time renormalization group equation let us start from the effective action at one loop level 
\begin{equation}
\Gamma[\phi]=S[\phi]+\frac{1}{2}\ln{\left[\det{S^{\left(2\right)}}\right]}=S[\phi]+\frac{1}{2}\mathrm{Tr}\ln{S^{\left(2\right)}}
\end{equation}
where $S^{(2)}$ is the Hessian of the theory and express the logarithm with the proper time representation 
\begin{equation}
\mathrm{Tr}\left[\ln{S^{\left(2\right)}}\right]=-\int_{0}^{\infty}{\frac{ds}{s}\mathrm{Tr} \left[e^{-sS^{\left(2\right)}}\right]}
\end{equation}
here the variable $s$ acts as a proper time for the quantum system. In particular we can relate it to the inverse of the momentum square of the fields. In this way the UV divergences of the field theory now are contained in the integral that has to be regularized introducing a cutoff $\Lambda$ which suppresses the contributions for small $s$, $s<\Lambda^{-2}$. Then to implement a Wilsonian coarse graining we introduce an IR cutoff $k$ which suppresses the contributions for large $s$, $s>\Lambda^{-2}$.  In this way 
\begin{equation}
\int_{0}^{\infty}ds\rightarrow\int_{1/\Lambda^2}^{1/k^2}ds
\end{equation}
this yields a standard sharp cutoff regularization, however it is more useful to introduce a more general class of propertime cutoffs by replacing 
\begin{equation}
\int_{0}^{\infty}ds\rightarrow\int_{0}^{\infty}f\left(s,k^2,\Lambda\right)ds
\end{equation}
The cutoff function $f\left(s,k^2,\Lambda\right)$ interpolates smoothly between $f\left(s,k^2,\Lambda\right)\approx 0$ for $s\gg k^{-2}$ and $f\left(s,k^2,\Lambda\right)\approx 1$ for $s\ll k^{-2}$. The first condition eliminates the UV singularity as $s\to 0$, the second ensures that the IR behavior remains unaffected by the introduction of the cutoff. Formally we require 
\begin{equation}
\lim\limits_{s\rightarrow+\infty}f\left(k\neq0,\mathrm{\Lambda},s\right)=0, \quad \quad \lim\limits_{s\rightarrow0}f\left(k=0,\mathrm{\Lambda},s\right)=1.
\end{equation}
Finally, since at the UV scale $\Lambda$ only the bare theory is present, the one-loop contribution must vanish so we demand 
\begin{equation}
\lim\limits_{k\rightarrow\mathrm{\Lambda}}f\left(k,\mathrm{\Lambda},s\right)=0. 
\end{equation}
Therefore we find
\begin{equation}
\Gamma_k\left[\phi\right]=S\left[\phi\right]-\frac{1}{2}\int_{0}^{\infty}f\left(s,k^2,\Lambda\right)\mathrm{Tr} \left[e^{-sS^{\left(2\right)}}\right]ds
\end{equation}
the functional renormalization group equation is obtained taking the scale derivative of this result. However, since the equation was derived from a one-loop approximation the flow of $\Gamma_k [\phi]$ will be approximate too. To overcome this difficulty a renormalization group improvement can be exploited: replace the classical Hessian $S^{(2)}$ with the quantum hessian $\Gamma_k^{\left(2\right)}$. In this way the running of $\Gamma_k [\phi]$ includes an infinite resummation of diagrams and therefore improves on the one-loop approximation. The improved proper time flow equation (PTFE) is
\begin{equation}
\label{PTFE}
k\partial_k\Gamma_k [\phi]=-\frac{1}{2}\int_{0}^{\infty}{\frac{ds}{s}k\frac{\partial f\left(s,k^2,\Lambda\right)}{\partial k}\mathrm{Tr}\left[e^{-s\Gamma_k^{\left(2\right)}}\right]}
\end{equation}
for suitable cutoff functions the scale derivative is independent of $\Lambda$, consequently the limit $\Lambda\to\infty$ can be taken.

For suitable choice of the cutoff function the integral in eq.(\ref{PTFE}) can be performed exactly. For our computation we use the following cutoff
\begin{equation}
\label{cutoff}
f\left(s,k^2,\Lambda\right)=\frac{\Gamma\left(n+1,sk^2\right)-\Gamma\left(n+1,s\Lambda^2\right)}{\Gamma\left(n+1\right)}
\end{equation}
here $\Gamma(\alpha,x)=\int_{x}^{\infty}{t^{\alpha-1}e^{-t}dt}$ is the incomplete Gamma function and $n$ is an arbitrary positive real number which controls the behavior of $f(s,k^2,\Lambda)$ in the interpolating region. The scale derivatives of eq.(\ref{cutoff}) is
\begin{equation}
\label{dcut}
   k\partial_kf\left(k,\Lambda,s\right)\equiv\rho\left(s,k^2\right)=-2\frac{\left(sk^2\right)^{n+1}}{\Gamma\left(n+1\right)}e^{-sk^2}
\end{equation}
which is independent of $\Lambda$. This cutoff has extensively used in literature \cite{Liao:1995nm, Liao:1994fp, Schaefer:2001cn, Schaefer:1999em, Schaefer:1997nd, Meyer:1999bz, Papp:1999he, Litim:2010tt, Bonanno:2012dg, Bonanno:2022edf, Bonanno:2023ghc, Bonanno:2023fij} and in particular the first for precision calculations of the critical exponents \cite{Bonanno:2000yp}.

\section{The Polyakov action and the conformal anomaly}
In this section we briefly review what the Polyakov action and the conformal anomaly are. Let us consider a scalar field theory in two dimensions on a curved background manifold $\mathcal{M}$ with a Riemannian metric $g$
\begin{equation}
\label{SF2}
S\left[\phi,g\right]=\int{d^2x\sqrt g\phi [-\Box]\phi}
\end{equation}
the classical energy-momentum tensor for this theory is given by 
\begin{equation}
T^{\mu\nu}=\frac{2}{\sqrt g}\frac{\delta S\left[\phi,g\right]}{\delta g_{\mu\nu}}=\frac{1}{2}g^{\mu\nu}\partial_\alpha\phi\partial^\alpha\phi-\partial^\mu\phi\partial^\nu\phi
\end{equation}
and it is easy to check that it is traceless. This is due to the fact that eq.(\ref{SF2}) is invariant under the conformal transformation $g_{\mu\nu}\to e^{\sigma}g_{\mu\nu}$.

In the quantum domain the quantization of eq.(\ref{SF2}) is obtained from the generating functional
\begin{equation}
Z[J,\phi]=\int{D_g \phi \exp{\left(-S[\phi,g]+\int{d^2 x \sqrt g J\phi}\right)}}
\end{equation}
where $J$ is a scalar source field. The measure can be defined from
\begin{equation}
\int{D_g\phi e^{-(\phi,\phi)_g}}=1, \quad \quad (\phi,\phi)_g=\frac{1}{2}\int{d^2 x \sqrt g \phi^2}
\end{equation}
and it is invariant under diffeomorphism but not invariant under a conformal transformation.

The expectation value of the energy-momentum tensor is found from a functional derivative of $Z[J=0,\phi]$ with respect to the metric. The result then can be rewritten in terms of the effective action:
\begin{equation}
\left\langle T^{\mu\nu}\right\rangle=-\frac{2}{\sqrt g}\frac{\delta\mathrm{\Gamma}\left[\phi,g\right]}{\delta g_{\mu\nu}}
\end{equation}
this is equal to the classical result but the classical action is replaced by the effective action.

From the symmetries of eq.(\ref{SF2}) we expect relations between the correlation functions which are described by Ward-Takahashi identities. However, in the case of a conformal transformation the effective action transforms as
\begin{equation}
\delta_\sigma \mathrm{\Gamma}\left[\phi,g\right]=\int{d^2 x \delta_\sigma g_{\mu\nu} \frac{\delta\mathrm{\Gamma}\left[\phi,g\right]}{\delta g_{\mu\nu}}}
=\int{d^2 x \sqrt g \delta\sigma \left\langle T_\mu^\mu\right\rangle}
\end{equation}
since the measure is not conformally invariant, we expect that $\delta_\sigma \mathrm{\Gamma}\left[\phi,g\right]\ne 0$, so $\left\langle T_\mu^\mu\right\rangle \ne 0$. Therefore there is an anomaly related to the conformal transformation. Using the one-loop approximation for the effective action we can prove that 
\begin{equation}
\left\langle T_\mu^\mu\right\rangle=-\frac{1}{24\pi}R
\end{equation}
the well-known the trace anomaly. 

Starting from this result Polyakov \cite{POLYAKOV1981207} in 1981 functionally integrating the conformal anomaly was able to find the exact effective action on a curved manifold in two dimensions
\begin{equation}
\label{PolAc}
\mathrm{\Gamma}\left[g\right]=\frac{1}{96\pi}\int{d^2x\sqrt g R\frac{1}{-\Box}R}
\end{equation}
where $R$ is the Ricci scalar. The effective action is a non-local action containing the inverse of the covariant Laplacian $\Box$. This inverse operator has to be meant as the two point correlation function evaluated at two different point in the manifold $\mathcal{M}$.

\section{The Polyakov action from the PTFE}
The general effective average action $\Gamma_k [\phi]$ interpolates smoothly between the full quantum effective action and the bare action \cite{book}. Now we investigate whether the PTFE is able to recover eq.(\ref{PolAc}). To that end let us consider a two-dimensional manifold $(g,\mathcal{M})$ with an effective average action of the following form: 
\begin{equation}
\label{ans}
\mathrm{\Gamma}_k\left[\phi, g\right]=\int{d^2x \sqrt g [a_k+Rb_k+Rc_k(\Box)R]}+O(R^3)+\mathrm{\Gamma}_\text{k, matter}\left[\phi,g\right]
\end{equation}
where $a_k$, $b_k$ are coupling constants of the theory and $c_k (\Box)$ is a function of the covariant Laplacian called a “form factor”. The goal is to obtain the explicit expression of $c_k (\Box)$ and evaluate at $k=0$. The action $\mathrm{\Gamma}_\text{k, matter}\left[\phi,g\right]$ is a matter action.

The explicit expressions for $a_k$, $b_k$, $c_k (\Box)$ depend on the specific dynamic fields propagating, accordingly to perform the computation of the right-hand side of eq.(\ref{PTFE}) we have to choose a specific field theory. We consider a theory of a massive scalar field non-minimally coupled to a graviton in $d=2$: 
\begin{equation}
\label{tmf}
    \mathrm{\Gamma}_k\left[\phi,g\right]=\int{d^2x\sqrt g\left[\frac{1}{16\pi G_k}\left(R-2\lambda_k\right)+Rc_k(\Box)R+\frac{1}{2}\phi\left(-\Box+m^2+\xi R\right)\phi\right]}
\end{equation}
in this action we are neglecting the running of $m$ and $\xi$ so the matter field is a classical field. A comparison to eq.(\ref{ans}) shows that the coupling constant $b_k$ is related to the Newtonian constant $G_k$ by $b_k=1/16\pi G_k$ whereas $a_k$ is related to the cosmological constant by $a_k=-\lambda_k/8\pi G_k$.

To compute the hessian of eq.(\ref{tmf}) we should decompose the metric splitting $g_{\mu\nu}$ in a background field $\bar{g}_{\mu\nu}$ and a fluctuation field $h_{\mu\nu}$, namely $g_{\mu\nu}=\bar{g}_{\mu\nu}+h_{\mu\nu}$, however in $d=2$ the Ricci scalar is a topological term so it can be neglected in the dynamic and $R c_k(\Box)R$ gives terms with more than two covariant derivatives in $h_{\mu\nu}$, which are of subleading order so we neglect this piece. For the remaining term $\phi R\phi$ we retain only the leading term coming from the background: $\phi^2 R=\phi^2 \bar{R}+O(h_{\mu\nu})$. Therefore, the hessian reads $\Gamma_k^{\left(2\right)}=-\Box+m^2+\xi \bar{R}$. Inserting into the proper time flow equations we get 
\begin{equation}\begin{split}
&k\partial_k\Gamma_k=-\frac{1}{2}\int_{0}^{\infty}{\frac{ds}{s}\rho\left(s,k^2\right)\mathrm{Tr}\left[e^{-s\left(-\Box+m^2+\xi \bar{R}\right)}\right]}=\\
&=-\frac{1}{2}\int_{0}^{\infty}{\frac{ds}{s}\rho\left(s,k^2\right)e^{-sm^2}\mathrm{Tr}\left[e^{-s\left(-\Box+\xi \bar{R}\right)}\right]}
\end{split}\end{equation}
from this result we can read the running of $\lambda_k=-8\pi G_k a_k$, $G_k=1/16\pi b_k$ as well as $c_k(\Box)$, as we will see in the next lines. To solve the equations then we need of a boundary condition. We assume that the bare action coincides with the matter action. This action sets the boundary conditions for $a_k$, $b_k$ and $c_k$ to solve the flow equations.

The trace can be evaluated using an expansion in powers of curvature invariants developed by Barvinsky and Vilkovisky \cite{Barvinsky1987BeyondTS,Barvinsky:1990up, Barvinsky:1990uq, Barvinsky:1993en}:
\begin{equation}\begin{split}
\label{NLHK}
&\mathrm{Tr}\left[e^{-s(-\Box\mathbf{1}+\mathbf{U})}\right]=\frac{1}{4\pi s}\int{d^2x\sqrt g\mathrm{tr}} \bigg\{\mathbf{1}-s\mathbf{U}+s\frac{R}{6}\mathbf{1}+\\
&+s^2\big[Rf_{R2d}\left(-s\Box\right)R+Rf_{RU}\left(-s\Box\right)\mathbf{U}+\mathbf{U}f_U\left(-s\Box\right)\mathbf{U}+\Omega_{\mu\nu}f_\Omega\left(-s\Box\right)\Omega^{\mu\nu}\big]+O\left(R^3\right)\bigg\} 
\end{split}\end{equation}
here $\mathbf{1}$ denotes the unit matrix of the space of fields on which $\Box$ acts and
\begin{equation}\begin{split}
&f_{R2d}\left(x\right)=\frac{1}{32}f\left(x\right)+\frac{1}{16x}\left[2f\left(x\right)-1\right]+\frac{3}{8x^2}\left[f\left(x\right)-1\right]\\
&f_{RU}\left(x\right)=-\frac{1}{4}f\left(x\right)-\frac{1}{2x}\left[f\left(x\right)-1\right],\quad\quad f_U\left(x\right)=\frac{1}{2x},\quad\quad f_\Omega\left(x\right)=-\frac{1}{2x}\left[f\left(x\right)-1\right]
\end{split}\end{equation}
where $f(x)$ is the structure factor 
\begin{equation}
f(x)=\int_{0}^{1}{d\alpha e^{-x\alpha\left(1-\alpha\right)}}.
\end{equation}
and $\Omega_{\mu\nu}$ is defined from $[D_\mu,D_\nu]\psi=\Omega_{\mu\nu} \psi$. Eq.(\ref{NLHK}) sums all contributions of the trace of the heat kernel in terms of a curvature expansion, we truncated this expansion at the second order in the curvature. In two dimensions the Ricci scalar is the only independent curvature invariant we can have due to $R_{\mu\nu}=\frac{1}{2}g_{\mu\nu}R$.

Performing the integrals, where in our case $U=\xi R\mathbf{1}$ and the trace of identity is simply one, using eq.(\ref{dcut}) and comparing to the scale derivative of eq.(\ref{ans}) we find
\begin{equation}
k\partial_k a_k=\frac{k^2}{4\pi n\left(1+\frac{m^2}{k^2}\right)^n},\quad \quad k\partial_k b_k=\frac{1-6\xi}{24\pi\left(1+\frac{m^2}{k^2}\right)^{1+n}}
\end{equation}
for the flow of coupling constants and
\begin{equation}\begin{split}
\label{betac}
    &{k\partial}_kc_k\left(y\right)=\frac{1-8\xi}{64\pi y\left(1+\frac{m^2}{k^2}\right)^{n+1}}-\frac{3k^2}{32\pi n y^2\left(1+\frac{m^2}{k^2}\right)^n}+\\
    &+\int_{0}^{1}\bigg[\frac{(n+1){(1-4\xi)}^2k^{2\left(n+1\right)}}{128\pi\left(k^2+m^2+\left(\alpha-1\right)\alpha y\right)^{n+2}}+\frac{(4\xi-1)k^{2\left(n+1\right)}}{32\pi y\left(k^2+m^2+\left(\alpha-1\right)\alpha y\right)^{n+1}}+\\
&+\frac{3k^{2\left(n+1\right)}}{32\pi n y^2\left(k^2+m^2+\left(\alpha-1\right)\alpha y\right)^n}\bigg]d\alpha
\end{split}\end{equation}
for the flow of the form factor, where $y=\Box$. 

Eq.(\ref{betac}) involves a complicated integral, to see the explicit form we fix $n$ and analyze the results for different values of $n$. In fig. \ref{betafig} the beta function for different values of $n$ is shown. In the plot \ref{betafig}(a) we show it as function of $k$ in unit of mass $m$ for fixed $y$ and $\xi$. The plots are similar, the beta functions are defined for all positive values of $k$ but as $n$ increases the maximum becomes larger and shifts. In the plot \ref{betafig}(b) we show the beta function at fixed $k$ versus $y$. The main feature is that as $n$ increases the effect of the cutoff function is to select a smaller and smaller momentum contribution and produce a sharp cut-off limit as $n\to\infty$.

\def\figsubcap#1{\par\noindent\centering\footnotesize(#1)}
\begin{figure}[t]%
\begin{center}
  \parbox{2.1in}{\includegraphics[width=1.1\linewidth]{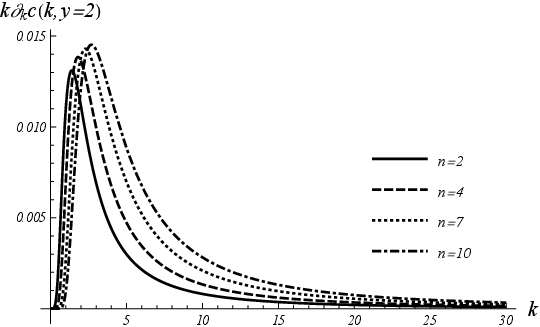}\figsubcap{a}}
  \hspace*{20pt}
  \parbox{2.1in}{\includegraphics[width=1.1\linewidth]{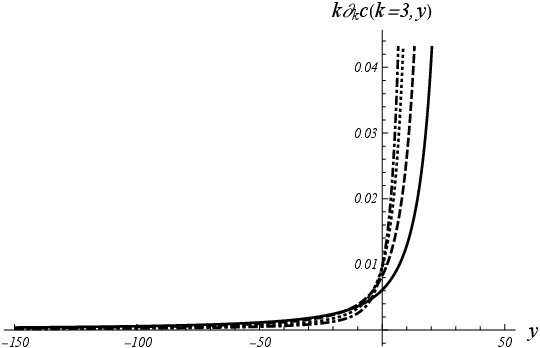}\figsubcap{b}}
  \caption{Beta function of the form factor  in unit of mass $m$ for different values of $n$. In (a) the beta function vs $k$ with $y=2$ and $\xi=1$. In (b) the beta function vs $y$ for $k=3$, $\xi=1$.}%
  \label{betafig}
\end{center}
\end{figure}

\subsection{The running of the coupling constants}
Integrating the beta functions for the coupling constants between an IR scale $k$ and an UV scale $\Lambda$ we get 
\begin{equation}\begin{split}
& a\left(k\right)-a\left(\Lambda\right)=-\frac{m^2\left(H_n+\ln{\left(\frac{m^2}{\Lambda^2}\right)}-1\right)}{8\pi}+\\
&+\frac{k^2\left(\frac{k}{m}\right)^{2n}\Gamma{\left(n\right)}{_2\tilde{F}}_1{\left(n,n+1;n+2;-\frac{k^2}{m^2}\right)}}{8\pi}-\frac{\Lambda^2}{8\pi n}+O\left(\frac{1}{\mathrm{\Lambda}}\right)\\
& b\left(k\right)-b\left(\Lambda\right)=\left(1-6\xi\right)\bigg[\frac{\left(\frac{k}{m}\right)^{2\left(n+1\right)}F_1{\left(n+1,n+1;n+2;-\frac{k^2}{m^2}\right)}}{48\pi\left(n+1\right)}+\\
&+\frac{H_n+\ln{\left(\frac{m^2}{\Lambda^2}\right)}}{48\pi}\bigg]+O\left(\frac{1}{\Lambda}\right)
\end{split}\end{equation}
where $H_n$ is the $n$-th harmonic number $H_n=\sum_{p=1}^{n}\frac{1}{p}$ and $_2\tilde{F}_1{\left(a;b;c;z\right)}$ is the regularized hypergeometric function $_2\tilde{F}_1{\left(a;b;c;z\right)}= _2 F_1{\left(a;b;c;z\right)}/\Gamma\left(c\right)$. The quantities $a\left(\Lambda\right)$ and $b\left(\Lambda\right)$ are free coefficients.

The coupling constants are divergent, in particular $a(k)$ has a quadratic and a logarithmic divergence, whereas $b(k)$ has only a logarithmic divergence. To remove these divergences we renormalize the theory setting
\begin{equation}
a\left(\Lambda\right)=\frac{\Lambda^2}{8\pi n}+m^2\log{\left(\frac{\mu^2}{\Lambda^2}\right)},\quad \quad b\left(\Lambda\right)=-\log{\left(\frac{\mu^2}{\Lambda^2}\right)}
\end{equation}
where we introduced a renormalization scale $\mu$. From the renormalized results we can take the limit $k=0$ to get the physical coupling constants 
\begin{equation}
a\left(0\right)=-\frac{m^2\ln{\left(\frac{m^2}{\mu^2}\right)}}{8\pi},\quad \quad b\left(0\right)=\left(1-6\xi\right)\frac{\ln{\left(\frac{m^2}{\mu^2}\right)}}{48\pi}
\end{equation}
they are trivial constants. In the massless limit $a(0)$ vanishes but $b(0)$ diverges unless the conformal limit $\xi=1/6$ is considered.

\subsection{The running of the form factor}
\begin{figure}[t!]
    \centering
    \includegraphics[width=0.8\linewidth]{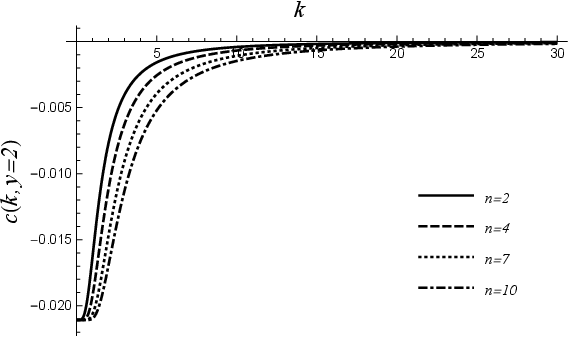}
    \caption{The form factor in unit of mass for different values of $n$. All form factors converge to a single value as $k\to0$. Here $\xi=1$.}
    \label{formc}
\end{figure}
Integrating the beta function of the form factor between an IR scale $k$ and the UV scale $\Lambda$, which we now take infinity, we get the form factor as a function of $k$. Fig. \ref{formc} shows the form factor for different values of $n$. They are defined for all positives $k$ and go to zero at infinity. The reason why this happens is that for dimensional reason at infinity only inverse powers of $\Lambda$ can appear, so the contribution at $k=\infty$ vanishes. This means that no renormalization is required. It can be also seen that as $n$ increases the slop increases too and for very large $n$ the form factor is a approximately an heaviside function. The most important feature of the plot is that altough the form factors show different behaviors for different values of $n$, as $k$ decreases they all converge to a single value at $k=0$. This is what we expect for the physical form factor. The result for the form factor at $k=0$ is:
\begin{equation}
\label{FormMass}
  c\left(k=0,y\right)=\frac{1}{96\pi y}+\frac{\xi}{8\pi y}+\frac{m^2}{16\pi y^2}+\frac{\tan^{-1}\left(\frac{1}{\sqrt{\frac{4m^2}{y}-1}}\right)}{\sqrt{\frac{4m^2}{y}-1}}\left(-\frac{\xi^2}{4\pi y}-\frac{m^2\xi}{2\pi y^2}-\frac{m^4}{4\pi y^3}\right)  
\end{equation}
In eq.(\ref{FormMass}) the first term is the Polyakov piece and we are interested in the way the Polyakov action is recovered. To that end we have to study the IR and the UV  limit of the result. The IR limit is defined from $m^2\gg y$, whereas the UV limit from $m^2\ll y$. To have a full understanding of the physical setting we discuss separately the case $\xi=0$ and $\xi\ne 0$. 

In the limit $\xi\rightarrow0$, we find
\begin{equation}
c\left(k=0,y\right)=\frac{1}{96\pi y}+\frac{m^2}{16\pi y^2}-\frac{\tan^{-1}\left(\frac{1}{\sqrt{\frac{4m^2}{y}-1}}\right)}{\sqrt{\frac{4m^2}{y}-1}}\frac{m^4}{4\pi y^3}
\end{equation}
we expect that Polyakov action is recovered as $m$ goes to zero. Indeed in the IR limit, $m^2\gg y$, we get
\begin{equation}
c\left(k=0,y\rightarrow0\right)=-\frac{1}{480\pi m^2}-\frac{y}{2240\pi m^4}-\frac{y^2}{10080\pi m^6}+O\left(y^3\right)
\end{equation}
the Polyakov term is absent, this means that the mass suppresses all quantum loops. On the contrary the Polyakov term comes up in the UV limit as the leading term of a massive expansion 
\begin{equation}
c\left(k=0,y\rightarrow+\infty\right)=\frac{1}{96\pi y}+\frac{m^2}{16\pi y^2}+\frac{m^4\ln{\left(-\frac{y}{m^2}\right)}}{8\pi y^3}+O\left(m^6\right)
\end{equation}
in the strict massless limit we remain only with the Polyakov term as expected. Actually there is a subtle point: we truncated at the second order in curvature. However, the conformal limit requires that the subsequent powers of the curvature vanish in the limit $m\rightarrow0$. This computation is beyond the scope of the present contribution which deals only with the second order form factors.

If $\xi\neq0$ the matter is slightly more complicated. In the IR limit $m^2\gg y$ we find again a local expansion
\begin{equation}
c\left(k=0,y\rightarrow0\right)=-\frac{30\xi^2+10\xi+1}{480\pi m^2}-\frac{(70\xi^2+28\xi+3)y}{6720\pi m^4}-\frac{(21\xi^2+9\xi+1)y^2}{10080\pi m^6}+O(y^3)
\end{equation}
and the Polyakov term is absent, in the UV limit we get
\begin{equation}\begin{split}
\label{UVlimit}
&c\left(k=0,y\rightarrow\infty\right)=\frac{1}{96\pi y}+\frac{\xi+\xi^2\log{\left(-\frac{y}{m^2}\right)}}{8\pi y}+\\
&+\frac{m^2\left(1-4\xi^2+4\xi^2\log{\left(-\frac{y}{m^2}\right)}+4\xi\log{\left(-\frac{y}{m^2}\right)}\right)}{16\pi y^2}+O\left(\frac{1}{y^3}\right)
\end{split}\end{equation}
the leading term is again the Polyakov piece but now it is corrected by a term that contains a logarithm and $\xi$. In the massless limit the non-leading terms go to zero but the leading term is singular due to the logarithm. This implies that you cannot have a massless limit if $\xi$ is different from zero and the pure Polyakov action cannot be recovered. In particular this implies that the limits $m\to0$ and $\xi\to0$ do not commute.

\section{Discussion}
The functional renormalization group is a promising tool to investigate non-perturbative phenomena and to develop the theory of quantum gravity. We saw that the proper time renormalization group equations are able to describe the flow of non-local terms. In particular, using a theory of a scalar field non-minimally coupled to gravity we showed how to derive the Polyakov action without integrating the conformal anomaly. The result is that only if $\xi=0$ the Polyakov action can be recovered in the massless limit.

The full effective action is a complicated non-local object. However, the IR and the UV limit of our results show two simple but different opposite behaviors. If the mass dominates the dynamic, the theory becomes local in the $\Box/m^2$ expansion. In the momenta space, this turns to be the standard decoupling limit of a particle in a loop. On the contrary if $\Box\gg m^2$ the theory is non-local, in particular it is a Taylor expansion in $m^2/\Box$. 

A limitation of our approach is that the computation of traces is quite involving and a general expression is not known. Only the trace of heat kernel operators or a function of a Laplace-type operator can be computed. This means that $\Gamma_k$ should be quadratic in the derivatives. The full $\Gamma_k$ coming from the integration of PTFE generates all sort of operators, including non-local terms, consequently to the full expression cannot be applied the heat kernel expansion. For this reason we considered only the flow of matter action.

\bibliographystyle{ws-procs961x669}
\bibliography{ws-pro-sample}

\end{document}